\documentclass[preprint,aps,pra,epsfig,preprintnumbers,superscriptaddress,floatfix]{revtex4}
\usepackage{graphics}
\usepackage[varg]{txfonts} 
\usepackage[latin1]{inputenc}
\bibstyle{prsty}
\begin{document}
\title{Feedback in a cavity QED system for control of quantum beats.}
\author{A. D. Cimmarusti}
\affiliation{Joint Quantum Institute, University of Maryland and NIST, College Park, MD 20742, USA}
\author{B. D. Patterson}
\affiliation{Joint Quantum Institute, University of Maryland and NIST, College Park, MD 20742, USA}
\author{C. A. Schroeder}
\affiliation{Joint Quantum Institute, University of Maryland and NIST, College Park, MD 20742, USA}
\author{L. A. Orozco}
\affiliation{Joint Quantum Institute, University of Maryland and NIST, College Park, MD 20742, USA}
\author{P. Barberis-Blostein}
\affiliation{IIMAS, Universidad Nacional Autonoma de M\'exico, M\'exico 04510, DF, M\'exico}
\author{H. J. Carmichael}
\affiliation{Department of Physics, University of Auckland, Private Bag 92019, Auckland, New Zealand}

\begin{abstract}
Conditional measurements on the undriven mode of a two-mode cavity QED system prepare a coherent superposition of ground states which generate quantum beats. The continuous system drive induces decoherence through the phase interruptions from Rayleigh scattering, which manifests as a decrease of the beat amplitude and an increase of the frequency of oscillation. We report recent experiments that implement a simple feedback mechanism to protect the quantum beat. We continuously drive the system until a photon is detected, heralding the presence of a coherent superposition. We then turn off the drive and let the superposition evolve in the dark, protecting it against decoherence. At a later time we reinstate the drive to measure the amplitude, phase, and frequency of the beats. The amplitude can increase by more than fifty percent, while the frequency is unchanged by the feedback.
\end{abstract}
\maketitle
\section{Introduction}
\label{intro}
The study of  open quantum systems through quantum trajectories \cite{carmichaelbookv1,carmichaelbookv2} has helped to elucidate the potential of quantum feedback \cite{wisemanbook}.  This is particularly important for the measurement of conditional intensities in quantum optical systems.
Our recent work in optical cavity QED uses such measurements to demonstrate the generation of ground state coherences by spontaneous emission \cite{norris10,norris12b}. Coherences arise from the internal structure of the rubidium atoms and its response to small magnetic fields. Superpositions of ground state magnetic sublevels Larmor precess and reveal their Zeeman shift in the frequency of a quantum beat.

During these investigations  we found that near-resonant Rayleigh scattering is responsible for an increase in the frequency of the beat note and a decrease in its amplitude \cite{norris12}. The present contribution reports on a first very simple protocol aimed at monitoring and reducing the decoherence due to this process. While measuring conditional intensities, i.e., the intensity correlation function $g^{(2)}(\tau)$, we turn off the system drive after the initial photon is detected, so that Rayleigh scattering is suppressed, although the coherence continues to precess.  We can then later return to a driven system in order to recover the amplitude and frequency of the oscillation. The time over which the quantum beats persist is sufficiently long that current experimental and theoretical tools allow us to implement the feedback and extract useful information to characterize the decoherence meachanism from the measurement \cite{deutsch10,barberis10,norris11}.

Our previous experiments on quantum feedback in optical cavity QED \cite{smith02,reiner04a} also worked with conditional intensities and weak drives. We controlled the oscillatory exchange of excitation between the atomic polarization and the cavity mode. The frequencies were five times higher than the decay rates of the system.  That previous protocol depended critically on the specific time of feedback application after a photon detection, as well as on the delay time of the electronics.
 
The protocol we report here relies on strong quantum feedback; we only require a single photon detection to obtain sufficient information to act on the system.  This is in contrast with two recent quantum control studies: cavity QED experiments with Rydberg atoms in superconducting cavities \cite{sayrin11,zhou12}, which use extensive calculations based on measurement outcomes to create and maintain a microwave Fock state; and quantum control of the full ground-state manifold of Cs \cite{smith06,merkel08,mischuck12}.

\section{System}
\label{sec:1}
We work with an optical cavity QED system in the intermediate coupling regime, where the single atom coupling, $g/2\pi=1.5\,{\rm MHz}$, is comparable to the decay rates of the cavity, $\kappa/2\pi=3.0\times10^6\,{\rm s}^{-1}$, and the atomic excited state, $\gamma/2\pi=6.07\times10^6\,{\rm s}^{-1}$. A detailed description of the apparatus can be found in Ref.~\cite{norris09a}. It consists of a $2\,{\rm mm}$ optical cavity, with $56\,\mu{\rm m}$ mode waist; the two orthogonal polarization modes of the cavity, which are resonant with the $D_2\,{}^{85}{\rm Rb}$ line, are degenerate to better than 20\% of the cavity half-width $\kappa$.

We collect ${}^{85}$Rb atoms in a magneto optical trap (MOT) placed $7.5\,{\rm cm}$ above the cavity, from which we direct them into the mode of the cavity through a $1.5\,{\rm mm}$ hole on the vacuum-located retro-reflecting optics (quarter wave plate and mirror) by an imbalance in the optical forces \cite{lu96}. The resulting continuous atomic beam has an average velocity between $15$ and $20\,{\rm ms}^{-1}$ depending on the MOT parameters. The apparatus provides the possibility of optically pumping the atoms into the $F=3, m=0$ ground state before they enter the cavity. A small magnetic field of 5 Gauss aligns the quantization axis with the polarization of the drive laser ($\pi$ polarization).

\begin{figure}
\begin{centering}
\resizebox{0.90\columnwidth}{!}{%
 \includegraphics{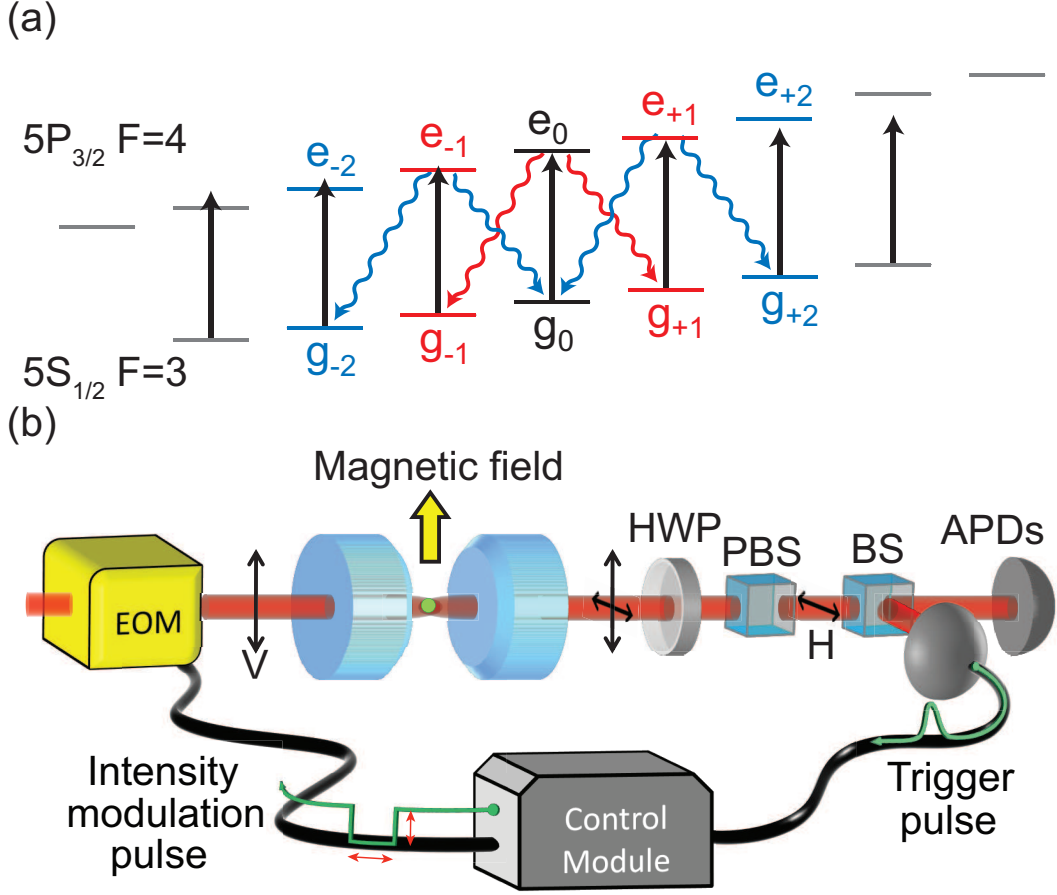} }
\caption{a) Atomic structure of $^{85}$Rb relevant to the experiment. b) Schematic of the apparatus after Ref.~\cite{cimmarusti12}.}
\label{fig:1}
\end{centering}
\end{figure}

The studied Zeeman coherences are established among the $m$ sublevels of the $F=3$ ground state in $^{85}$Rb [see Fig.~\ref{fig:1}(a)]. They are only visible in the conditional intensity \cite{norris10} and have their origin in a realization of the quantum eraser suggested by Zajonc \cite{zajonc83}. The observation of quantum beats in the conditional intensity---measured through the second-order correlation function---have a long history \cite{aspect84}, and spectroscopic methods relying on coherences are widely used across atomic, molecular, and condensed matter physics to probe the structure of materials (see for example Ref.~\cite{alexandrovbook}). The optical cavity provides the pair of orthogonal modes required to perform the experiment. We drive on the V mode and detect on the H mode. The V mode corresponds to $\pi$ polarization while the H mode is a linear combination of the two available $\sigma$ polarizations. Our selection of polarizations at the output ensure that we only observe light that has $\sigma$ components; their origin is spontaneous emission accompanied by a consequent change in the $m$ sublevel.

Figure \ref{fig:1}(b) shows a schematic of the experimental apparatus. The detection of a photon generates an electronic pulse which changes the amplitude of the laser drive for a pre-set amount of time. An electro-optic modulator (EOM) sets the drive intensity ($\pi$ polarized), which it can turn off completely or partially depending on the requirements of the study. Light exits the cavity containing both $\pi$ and $\sigma$ components. It passes through a half-wave plate (HWP), a polarizing beam splitter (PBS), and a beam splitter (BS), which direct photons onto a pair of avalanche photodiodes (APDs). A time stamp unit in a computer records the arrival times of the photo-pulses from the APDs  to obtain the conditional intensity, either $g^{(2)}(\tau)$ or other correlations. By selecting the angle at the HWP we may perform either a pure intensity measurement or a homodyne measurement of the conditional field. Appropriate electronics copy and shape the photopulses to drive the EOM and change the intensity \cite{cimmarusti12}.

Reference \cite{norris12} explains how the measured conditional oscillations suffer from decoherence which affects both their amplitude and frequency. This decoherence originates from random quantum jumps \cite{carmichael93book} caused by quasi-resonant Rayleigh scattering that interrupt the oscillation of the atomic dipole. The scattering occurs in an unfamiliar regime where the excitation lies within the linewidth of the excited state; the multilevel atomic structure allows coherence to persist over many scattering events. Our apparatus does not allow direct observation of the jumps, but their random nature induces a change in the measured amplitude of the quantum beat, and their accumulation at a mean rate proportional to time induces a frequency shift.  Observations in ion traps \cite{uys10} encounter related effects with far off-resonance excitation, and again within a multilevel atomic structure. Both observations (in ions and ours) make use of coherent excitation. The early days of optical pumping saw Cohen Tannoudji report a similar process occurring under incoherent excitation \cite{cohen62}.

\section{Model for frequency shifts and decoherence of quantum beats}
We present next the basic findings of Ref.~\cite{norris12} that relate the frequency shift and loss of amplitude to undetected Rayleigh scattering. The  analysis is valid for a single four-level atom---two ground states and two excited states---with constant coupling strength to a weak coherent drive. If we neglect $\sigma$ spontaneous emissions, it serves as closed four-level model. In practice it is an approximation to a 4-level submanifold that shuttles backward and forward in response to $\sigma$ spontaneous emission within the larger angular momentum manifold of our experimental system [see Fig.~\ref{fig:1}(a)]  as we track the coherence through many scattering events.  It neglects differences in detunings and Clebsh Gordon coefficients, but works rather well because we consider time scales that are short compared with the long-time optical pumping limit. Using quantum trajectories \cite{carmichael93book,plenio98}, we follow the evolution of a ground-state coherence, $|\psi_g\rangle =(|g_-\rangle+|g_+\rangle)/\sqrt2$, initially created between $m=\pm1$ Zeeman sublevels.

The evolution of the ground-state coherence acquires an additional relative phase advance (over and above that due to Larmor precession) and an additional change of relative amplitude due to backaction. Both amplitude and phase advance in discrete jumps when photons are scattered. So far as the amplitude is concerned, the component of the superposition with smaller (larger) scattering rate shrinks (grows) relative to the other. In between these events there is a reversal: the component of the superposition with smaller (larger) scattering rate grows (shrinks) relative to the other (a null-measurement result). As the Rayleigh scattering proceeds, one scattered photon after another, this evolution through quantum jumps is compounded.

When the rates of scattering on the two components of the superposition are unequal, the quantum trajectory effects a measurement of the spin state. The discrete amplitude change at each jump and the evolution between jumps compete, such that a fluctuation in the number of photons scattered makes the amplitude of one component overcome the other. An excess (deficit) of scattering brings about localization onto the spin state with larger (smaller) scattering rate. This dynamic corresponds to the usual case of projective state measurement through Rayleigh scattering.

When the system is in the weak drive regime (amplitude $\alpha$,  Rabi frequency  g$\alpha$) such that the number of photons in the cavity is smaller than the saturation photon number, and the drive is detuned by $\Delta$ (total detuning including Zeeman shifts), there is a new contribution, $-2\Delta_{jump}$, to the frequency of the beat, with
\begin{equation}
\Delta_{jump} =2g^2|\alpha|^2\Delta\left(\frac{\gamma/2}{(\gamma/2)^2+\Delta^2}\right)^2.
\label{jumpshift}
\end{equation}
This shift comes from the mean rate of phase accumulation from quantum jumps.  It is opposite in sign to the usual AC Stark shift brought about by the same driving laser, and larger in magnitude by a factor of two when the detuning is small ($\Delta\ll\gamma/2$). Observation of a frequency shift from quantum jumps requires at least the four-level structure, since jumps bring only an unobservable overall phase change to the state of a two-level atom under Rayleigh scattering.

There is also an added damping rate $\Gamma_{decoh}$, which decoheres the quantum beats, with
\begin{equation}
\Gamma_{decoh} = 2g^2|\alpha|^2\gamma\left(\frac{\Delta}{(\gamma/2)^2+\Delta^2}\right)^2.
\label{jumpwidth}
\end{equation}
The decoherence comes from diffusion of the phase of the ground-state superposition, a processes that accompanies the net phase drift which constitutes the frequency shift.  Both aspects are expected from the stochastic nature of the jump process.  Noting that $\Gamma_{decoh}/2\Delta_{jump}=2\Delta/\gamma$, we may speak of a well-resolved frequency shift when $\Delta\ll\gamma/2$.  If this condition is not met, a single jump can result in a phase change of $\pi$, which washes out the beat after a single jump over multiple measurements. This is the case explored by Uys {\it et al.} \cite{uys10}. Note that both the frequency shift [Eq.~(\ref{jumpshift})] and the damping [Eq.~(\ref{jumpwidth})] are linear in the number of intracavity drive photons $|\alpha|^2$.

Our experimental system brings many additional complications.  They include optical pumping within the full level structure of $^{85}$Rb, the finite interaction time and inhomogeneous coupling strengths associated with an atomic beam, (weak) many atom and cavity lifetime effects, and saturation at large driving strengths.  For these reasons we implement the theory as a full Monte Carlo simulation when making quantitative comparison with our measurements \cite{norris10}.

\begin{figure}
\begin{centering}
\resizebox{0.90\columnwidth}{!}{
 \includegraphics{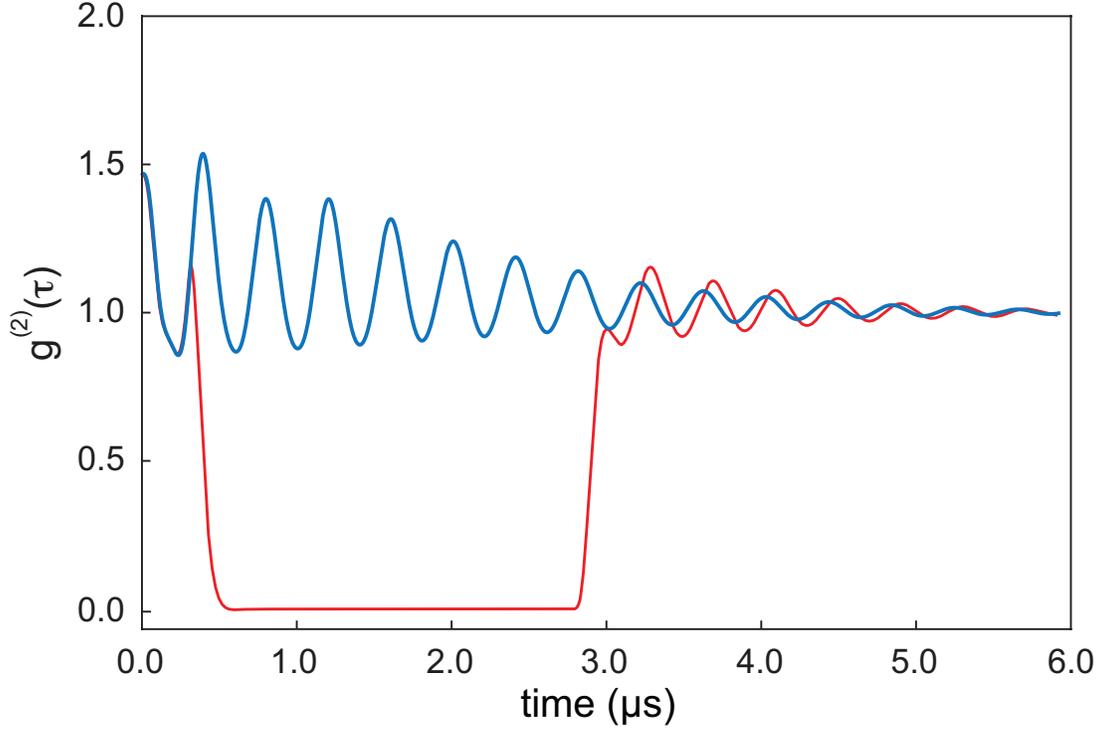} }
\caption{Results from a full Monte Carlo simulation showing the conditional intensity, $g^{(2)}(\tau)$, without feedback (blue) and with feedback (red).}
\label{fig:2}
\end{centering}
\end{figure}

The theoretical treatment starts out with a numerical simulation of the experiment in the absence of feedback to determine the best parameters for the effective number of atoms, number of photons in the driven mode, average atomic velocity, and the angle between the atomic beam and cavity axis. The simulation takes into account as many of the experimental realities as possible. It also includes a determination of the amount of drive light that passes the HWP after the cavity, effectively adding a homodyne measurement to the simple scheme described above. Homodyne measurement yields an oscillation at the Larmor precession frequency rather than at twice that frequency. The data and simulation in this paper follow that approach to observe the Larmor precession directly. The result of a simulation based on the experimentally measured correlation functions in Ref.~\cite{cimmarusti12} is presented in  Fig.~\ref{fig:2}. The optimal parameters used in this example are an effective atom number $N_{\rm eff}=0.55$, a HWP rotation of $1.2^\circ$, a mean atomic speed $v_{\rm p}=13.5\mkern2mu{\rm ms}^{-1}$, a deviation of the atomic beam from perpendicular to the cavity axis $\theta=0.017\mkern2mu{\rm rad}$, and mean number of photons in the driven cavity mode $n=1.21$.

The qualitative features of a loss of amplitude and a phase shift are clear in Fig.~\ref{fig:2}, where the trace in blue shows the Larmor precession without feedback and in red with the drive turned off for about 2.5 $\mu$s. The revived oscillations have a larger amplitude, and there is a clear phase shift when the coherence evolves without Rayleigh scattering. The simulation also shows additional structure in the early part of the oscillation (blue trace). This is related to the standing wave structure of the optical cavity and the specific paths followed by the atoms as they cross the mode.

\section{Results}
\begin{figure}
\begin{centering}
\resizebox{0.90\columnwidth}{!}
{ \includegraphics{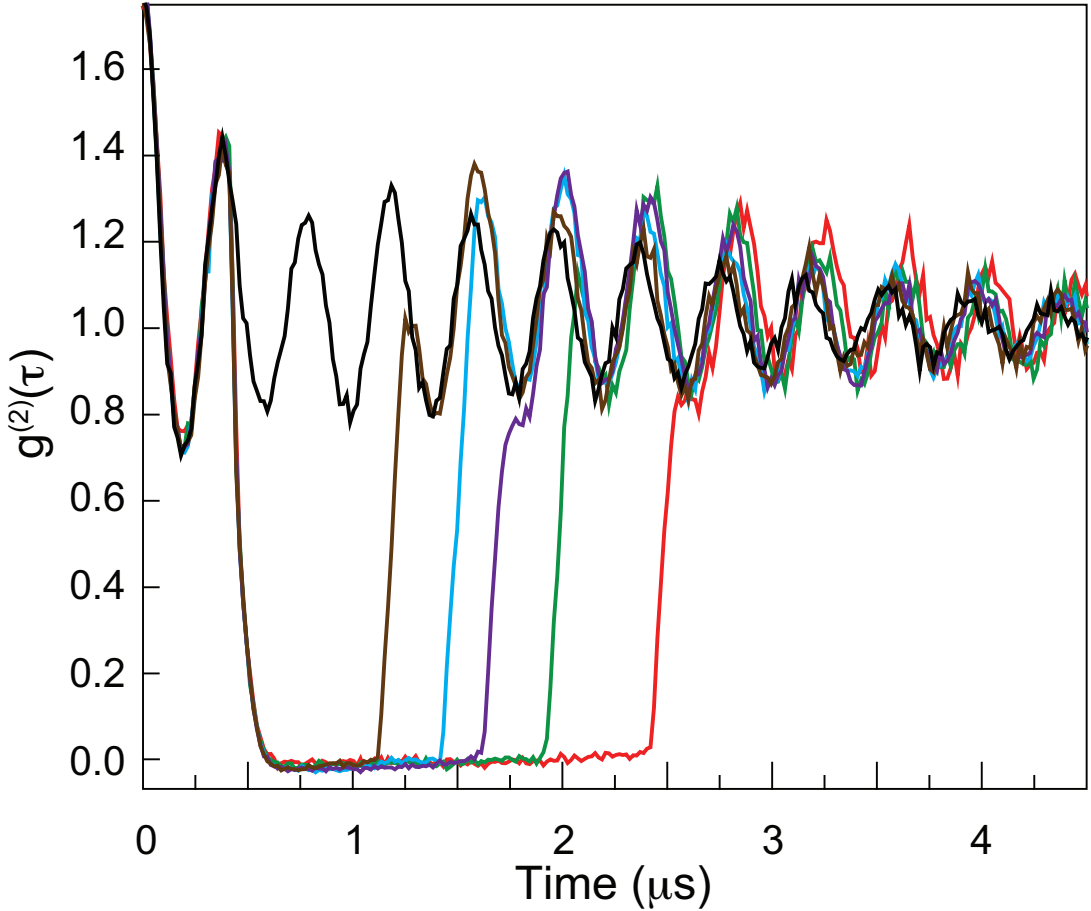} }
\caption{Measured conditional intensity, g$^{(2)}(\tau)$, with various feedback pulse lengths.}
\label{fig:3}
\end{centering}
\end{figure}

Figure \ref{fig:3} presents a typical set of measurements where we operate the system with a magnetic field close to 5 Gauss and atomic flux that keeps approximately 15 atoms inside the cavity mode to create an effective number of maximally coupled atoms around $N_{\rm eff}=2$. We record the conditional intensity of the undriven mode (H polarization) with our correlator after mixing a small amount of drive light (HWP angle about 1.2 degrees) to effect a measurement of the amplitude of the field. This allows us to look at the Larmor precession directly. We keep the drive at or near one photon in the cavity in steady state, while we collect data for a few minutes. After the detection of a photon we turn the light off, then, after a time that varies in the different traces, turn the drive back on. The results show clear changes in the amplitude and the phase (frequency) of the oscillation as a function of time with the drive turned off.

\begin{figure}
\begin{centering}
\resizebox{0.90\columnwidth}{!}{
 \includegraphics{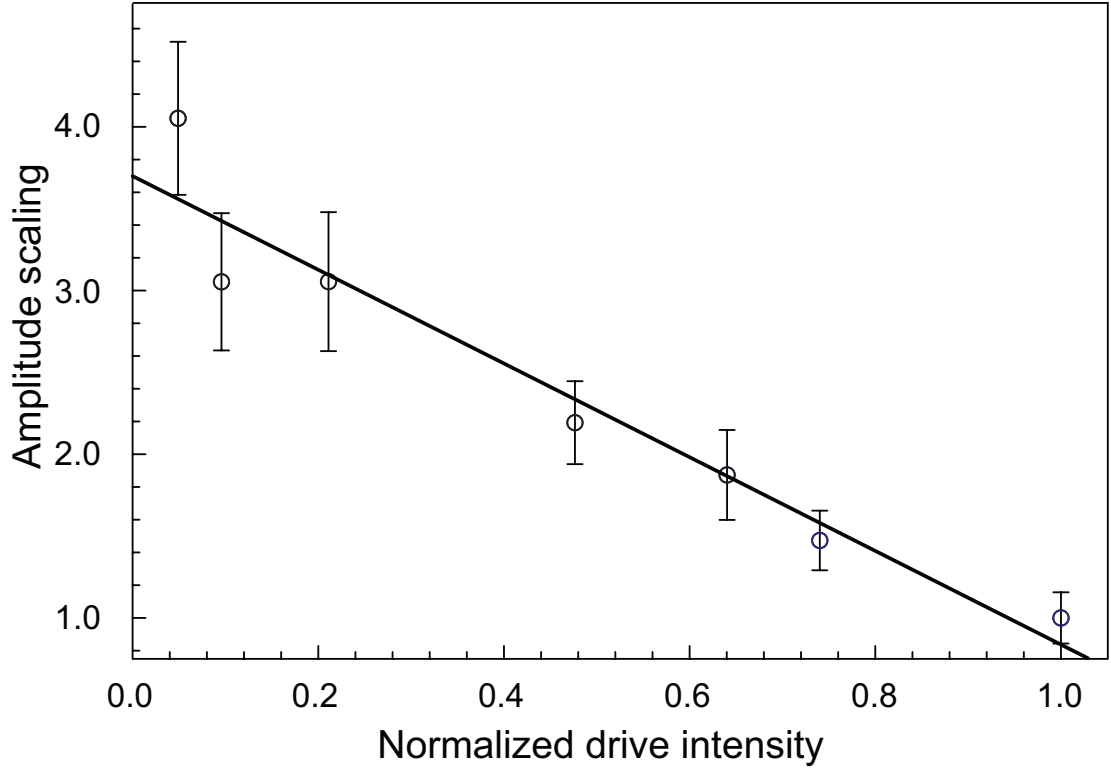} }
\caption{Amplitude of the recovered oscillation as a function of normalized feedback intensity after $3\,\mu{\rm s}$ of lower drive.}
\label{fig:4}
\end{centering}
\end{figure}

\begin{figure}
\begin{centering}
\resizebox{0.90\columnwidth}{!}{%
 \includegraphics{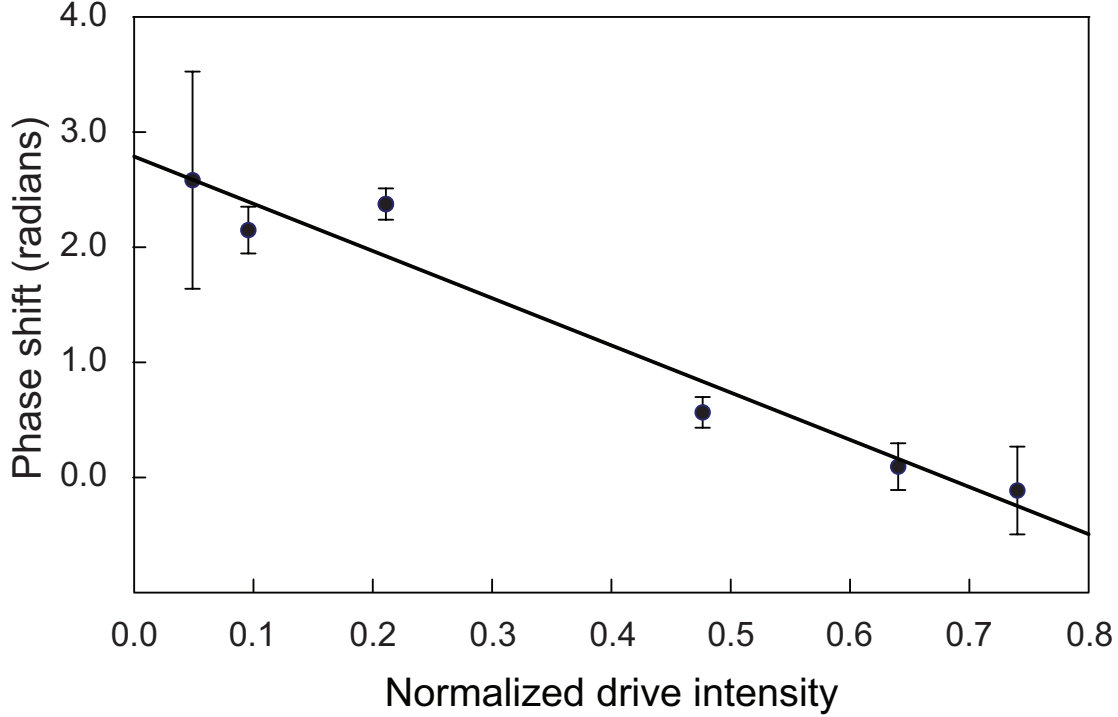} }
\caption{Phase shift as a function of normalized feedback intensity after $3\,\mu{\rm s}$ of lower drive.}
\label{fig:5}       
\end{centering}
\end{figure}

In Figs.~\ref{fig:4} and \ref{fig:5} we proceed to study the variation of the beat amplitude and frequency shift in a different way. We fix the
feedback pulse width at 3 $\mu$s and vary the amplitude of the drive applied during a partial turn off phase. This allows us to systematically look at the size of the oscillation and the phase shift after the light is turned back on. When we extract the data from the conditional intensity, we focus on the oscillations immediately following the return of the drive, assuming no significant decoherence or phase accumulation after the drive returns. Our assumption is supported by the very small change in phase and amplitude observed for the shortest pulse in Fig.~\ref{fig:3}.

Figure \ref{fig:4} shows how the amplitude of the recovered oscillation scales as a function of normalized drive intensity (normalized against the no feedback case).  If we completely turn the drive off, we obtain the largest amplitude, as observed both in the simulation of the experiment in Fig.~\ref{fig:2} and measurement results in Fig.~\ref{fig:3}. With the drive turned off for $3.0\,\mu{\rm s}$, the largest amplitude we observe is more than a factor of three greater than in the continuously driven case.

Figure~\ref{fig:5} shows the phase shift as a function of normalized intensity. For the smallest drive (about 5\%) the phase shift differs by more than 2 radians compared to the continuously driven case. The effective frequency shift is of the order of $100\,{\rm kHz}$ per drive photon in the cavity. We do not attempt to make a comparison with the prediction of the simple theory above of a linear dependence on photon number.  The continuous lines are merely to help guide the eye. A longer manuscript presents a direct comparison between theory and experiment with excellent results \cite{cimmarusti12}.

\section{Conclusions}
The subtle effects of Rayleigh scattering on the ground state quantum beats observed by us in previous work is possible to correct. We have demonstrated this by implementing a simple quantum feedback procedure following the detection of the photon that heralds the creation of the ground-state coherence. We simply turn off the drive and let the coherence evolve in the dark. In this way we avoid the quantum jumps that, although of small enough affect individually, occur sufficiently frequently to produce measurable frequency shifts and faster decoherence.

\section{Acknowledgments}
We wish to thank D. G. Norris for his continued interest in this project. Work supported by NSF of the USA and CONACYT, M{\'e}xico.


\end{document}